\documentclass[aps,pra,groupedaddress,twocolumn,showpacs]{revtex4}

\usepackage{amssymb}
\usepackage{graphicx}
\usepackage{pstricks}
\usepackage{amsmath}

\newcommand{\ket}[1]{\ensuremath{|#1\rangle}}

\newcommand{\TLmatrix}[2]{\begin{array}{ccc} 0 & #1 & 0 \\ #1 & 0 &  #2 \\ 0 &  #2 & 0 \end{array}}

\begin{document}
\title{Filtering of matter wave vibrational states via spatial adiabatic passage}

\author{Yu. Loiko,$^{1,2}$ V. Ahufinger,$^{1,3}$ R. Corbal\'{a}n,$^{1}$ G. Birkl,$^{4}$ and J. Mompart$^{1}$}

\affiliation{$^1$Departament de F\'{\i}sica, 
Universitat Aut\`{o}noma de Barcelona, E-08193 Bellaterra, Spain } 

\affiliation{$^2$Institute of Physics,
National Academy of Sciences of Belarus, Nezalezhnasty Ave. 68, 220072 Minsk, Belarus}

\affiliation{$^3$ICREA - Instituci\'{o} Catalana de Recerca i Estudis
Avan\c{c}ats, Llu\'is Companys 23, E-08010 Barcelona, Spain}
 
\affiliation{$^4$Institut f\"ur Angewandte Physik, Technische Universit\"at Darmstadt,
Schlossgartenstr 7, D-64289 Darmstadt, Germany}

\begin{abstract}
We discuss the filtering of the vibrational states of a cold atom in an optical trap, 
by chaining this trap with two empty ones and controlling adiabatically the tunneling. 
Matter wave filtering is performed by selectively transferring the population of the 
highest populated vibrational state to the most distant trap 
while the population of the rest of the states remains in the initial trap. 
Analytical conditions for two-state filtering are derived and then applied 
to an arbitrary number of populated bound states. 
Realistic numerical simulations close to state-of-the-art experimental arrangements are 
performed by modeling the triple well with time dependent P\"oschl--Teller potentials. 
In addition to filtering of vibrational states, we discuss applications 
for quantum tomography of the initial population distribution and 
engineering of atomic Fock states that, eventually,
could be used for tunneling assisted evaporative cooling.
\end{abstract}

\date{\today }
\maketitle

\section{\label{Introduction}Introduction}

Ultracold atoms trapped in optical potentials \cite{saop,grangier,latestbirkl}, e.g., microtrap arrays or optical lattices, have attracted considerable attention since they fulfill all the basic requirements for quantum information processing \cite{div}. In fact, neutral atoms in dipole trap arrays with short-range interactions such as s-wave scattering of bosons \cite{swave} or state selective long range interactions such as dipole-dipole interactions \cite{longrange}, do not experience intrinsic limitations in their scalability. Thus, a quantum register of about a hundred qubits has been reported recently in a two-dimensional (2D) optical microtrap array \cite{latestbirkl}. 2D optical microtrap arrays present two characteristic features: (i) the simplicity to achieve single-site addressing since the trap separation distances can range from
single ${\rm \mu m}$ up to $\sim 100\, {\rm \mu m}$, in any case being larger than the resolution limit; and (ii) the freedom to move independently sets of traps or even individual traps to control the tunnelling.    
However, cooling a single atom down to the lowest vibrational state of an optical microtrap array is still a challenging issue and, for some physical realizations, represents one of main experimental limitations to perform quantum computations with optical microtraps. Thus, developping novel techniques to determine and, eventually, to engineer the population distribution in optical microtraps is a focus of present research \cite{latestbirkl}.
In this context, we here propose to make use of 
the spatial adiabatic passage technique \cite{matterwaveSTIRAP} to achieve this goal. 

Spatial adiabatic passage consists in adiabatically following a spatial dark state whose spatial profile is determined by the tunneling interaction between neighboring traps \cite{matterwaveSTIRAP} and, in fact, is the matter wave analog of the well known quantum optical Stimulated Raman Adiabatic Passage (STIRAP) technique \cite{STIRAP}. Here, we will take profit of the fact that tunneling rates between traps strongly depend on the vibrational state under consideration to perform state selective adiabatic passage leading to (i) filtering of vibration states, (ii) quantum tomography of the initial population distribution, and (iii) engineering of atomic Fock states that, eventually, could be used for tunneling assisted evaporative cooling.

The article is organized as follows. In Section II we introduce the physical model consisting in a single atom in three identical P\"{o}schl--Teller type potentials \cite{PoschlTeller} with time varying position of the trap centers. The filtering protocol is presented in Section III and analytical conditions for two-state filtering are derived and compared to numerical simulations. Section IV focuses on the application of the filtering protocol to multiple-states for quantum tomography and quantum engineering of Fock states. Finally, the conclusions are summarized in Section V.

\section{\label{Model}Model}

\begin{figure}[t]
\begin{center}
\includegraphics[scale=1.0,clip=true,angle=0]{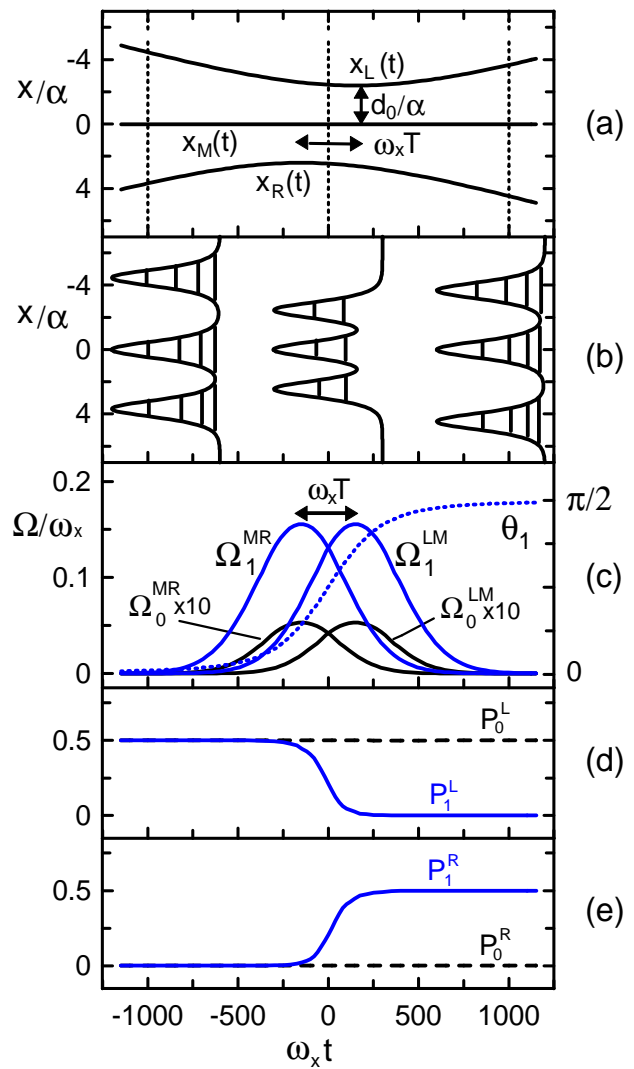}
\end{center}
\caption{
(Color online) 
(a) Temporal variation of the positions of the PT traps centers, where $d_0$ is the minimum trap separation and $T$ is the time delay between the two approaching sequences. (b) Illustration of the spatial profile of the PT potentials at the three different times corresponding to the vertical lines in (a). 
Each isolated PT trap has depth of ${V}_{0}=-20 \hbar \omega_x $ and supports only 4 vibrational states. (c) Temporal variation of the tunneling rate $\Omega_n^{LM}$ ($\Omega_n^{MR}$) between left and middle (middle and right) traps for the ground ($n=0$) and first ($n=1$) excited states. $\theta_1$ is the mixing angle for the first excited states. Temporal variation of the population of the ground and first excited states of the left (d) and right (e) traps assuming the following initial distribution $P_0^{L}(t=0)=P_1^{L}(t=0)=1/2$, respectively.}
\label{f1v1}
\end{figure}

We study the dynamics of a single cold neutral atom of mass $m$ in a 1D triple well potential, see Figs.~\ref{f1v1}(a) and (b),
described by the Schr\"{o}dinger equation: 
\begin{equation}
i \hbar\frac{\partial }{\partial t}\psi (x,t)=\left[-\frac{\hbar^2}{2m}\frac{\partial^2}{\partial x^2}+{V}\left( 
x,t\right)\right ] \psi (x,t), \label{Schrodinger_eq}
\end{equation}
where the wells are modeled by three identical P\"{o}schl--Teller (PT) type potentials \cite{PoschlTeller}:
\begin{equation}
V\left( x,t\right) =\sum_{i=L,M,R}V_{0i} {\rm sech}^{2}\left[  \sqrt{2}\left( x-x_{i}\left( t\right) \right) / {\alpha} \right].
\label{Eq_MWT_VxPoschlTeller}
\end{equation}
$V_{0i}$ is the potential depth for the $i$-th trap, and $x_{i}\left(t\right)$ defines the position of its center at time $t$. 
$2 \alpha$ is the width of the PT potential and
$\omega _{x} = \hbar /m \alpha^{2}$ is the trapping frequency. 
Note that we assume here PT potentials since they can be used to model very accurately Gaussian potentials
(see Appendix A),
obtained with dipole traps built-up by focusing a laser beam.
In this case the parameter $\alpha$ corresponds to the waist $w_{0}$ of the Gaussian light beam. In addition, 
PT potentials provide analytical expressions for their energy eigenvalues
and eigenstates, see Sec.III B. 
Throughout the paper, we will use dimensionless units for time, 
$t\omega_{x}$, space, $x/\alpha$, and potential amplitude, 
$ V_0 /\hbar \omega_{x}$. 
Initially, at $t_{ini}$, we will assume that the neutral atom is 
distributed among the vibrational states of the left trap, 
while the other two traps are empty.

\section{\label{Two_state_filtering}Two-state filtering }

\subsection{\label{Basic_idea} Basic idea}

Our proposal for the filtering of vibrational states is based 
on an adiabatic transport process \cite{matterwaveSTIRAP} 
between the two extreme traps 
that resembles the well known quantum optical Stimulated Raman Adiabatic Passage (STIRAP) technique \cite{STIRAP}. 
Although we will accurately investigate the filtering of vibrational states by numerically solving the Schr\"odinger Eq. (1), 
for simplicity, in the following lines we will illustrate the basics of our proposal by restricting the 
dynamics into the Hilbert space spanned by the ground and the first excited vibrational states of each trap. 
In this finite basis, the Hamiltonian of the system can be roughly approximated to $H=H_0\oplus H_1$ with:
\begin{eqnarray}
H_{n} = \hbar \left( \TLmatrix{{\Omega}^{LM}_{n}(t)}{{\Omega}^{MR}_{n}(t)} \right),\label{Ham_n}
\end{eqnarray}
where $\Omega^{ij}_{n}$ is the tunneling rate between two adjacent traps $i$ and $j$ with $i,j=L,M,R$ denoting left, middle and right, respectively and $n=0,1$ refers 
to the ground or the first excited vibrational state, respectively. 
Note that we have split Hamiltonian $H$ into the direct sum of $H_0$ and $H_1$ and, therefore, 
we have assumed that the energy separation between the ground and first vibrational state of each trap 
is large enough to avoid crossed tunneling between ground and excited states of different traps.
This approximation could fail for vibrational states 
close to the continuum, such as the vibrational analogues of Rydberg states, where
the energy spacing between different levels becomes relatively small. The latter scenario is out of the scope of this paper.

After diagonalization of the two Hamiltonians given in Eq.~(3), one ends up, in particular, 
with two energy eigenstates that only involve vibrational states of the two extreme traps:
\begin{eqnarray}
\ket{D_n ({\theta_n})}&=&\cos\theta_n \ket{n}_L - \sin\theta_n \ket{n}_R \quad {\rm with} \, n=0,1\,\,
\label{eqn:darkstates}
\end{eqnarray} 
where the mixing angle, $\theta_n$, is defined as $\tan \theta_n \equiv \Omega_n^{LM} / \Omega_n^{MR}$.
States $\ket{D_n ({\theta_n})}$ are known as spatial dark states \cite{matterwaveSTIRAP}.
State-selective adiabatic passage of matter waves between the two extreme traps 
will consist in adiabatically following one of the two energy eigenstates (4), 
typically the one with highest energy, by the smooth variation of the tunneling rates, 
while for the other one the transport process is inhibited. 

\subsection{\label{PTTR} P\"oschl--Teller tunneling rates}

For a single PT potential (\ref{Eq_MWT_VxPoschlTeller})
of depth $V_{0}=-s\left( s+1\right)$ ($s>0$)
there is an analytical solution for their energy eigenstates in terms of 
the associated Legendre $P_{s}^{\left(n-s\right) }$ polynomials as follows:
\begin{equation}
\phi _{n,s}\left(x\right) =N_{n,s}  P_{s}^{\left( n-s\right) }
\left[\tanh ( \sqrt{2} \left(x - x_{i}\right) / \alpha ) \right]
\label{Eq_MWT_PoschlTeller_nL_Legendre}
\end{equation}
with normalization constant:
\begin{equation}
N_{n,s} =\frac{2^{1/4}}{\alpha^{1/2}} 
\sqrt{\left(s-n\right) \frac{\Gamma \left(
2s-n+1\right) }{\Gamma \left( n+1\right) }}~,
\label{Eq_MWT_PoschlTeller_nL_Legendre_norm} 
\end{equation}
where 
$s =\sqrt{\left| V_{0} \right|+1/4}-1/2$, 
$n =0,1,\dots, N_{max}$ numerates the bound states with $N_{max}$ being the integer part of $s$, and 
$E_{n,s}/\hbar \omega_x =-\left( s-n\right) ^{2}$ gives the energy of the $n$-th state. $\Gamma$ is the Gamma function. 

The tunneling rates $\Omega_{n,s}$ between two identical PT potentials can be found by determining the energy difference between 
the symmetric $\phi _{n,s}^{+}$ and antisymmetric $\phi _{n,s}^{-}$ eigenstates 
namely $\Omega _{n,s} =
\left\langle \phi_{n,s}^{+} \left| H \right| \phi _{n,s}^{+}\right\rangle 
- \left\langle \phi _{n,s}^{-} \left| H \right| \phi_{n,s}^{-}\right\rangle =
E_{n,s}^{+} -E_{n,s}^{ -}$. We take $\phi^{\pm }_{n,s}=
\left( \phi _{n,s}^i \pm \phi_{n,s}^j \right) /\sqrt{2}$ where $\phi _{n,s}^{i,j}$ corresponds, see Eq. (5), to the localized state $\phi _{n,s}$ of either the trap $i$ or its neighbour $j$. For $n=0$, the Gram-Smith orthonormalization procedure (see Appendix \ref{AppendixB}) provides very accurate analytical expressions for the ground state tunneling rate, $\Omega_{0,s}$. 
For the exited states, only approximate analytical solutions are possible. The Holstein--Herring method, see \cite{1952Holstein56JPC832,1962Herring34RMP631}, yields for two identical traps:
\begin{equation}
\Omega_{n,s} (d)
= E_{ n,s }^{+}-E_{ n,s }^{-} 
= \left. \frac{-\nabla \left( \phi _{n,s}^{ i }\left( x\right) \right) ^{2}}{1-2\int_{x}^{\infty }\left( \phi _{n,s}^{ i }\left( x'\right) \right) ^{2}dx'}\right| _{x=d/2}
~,  \label{Eq_dE_Holstein_Herring}
\end{equation}
where $d$ is the distance between the trap centers.

At large distances between the traps, the denominator in (\ref{Eq_dE_Holstein_Herring})
rapidly approaches unity and the main behavior of the tunneling rate
$\Omega _{n,s}\left( d \right)$ is given predominantly by the numerator,
for which by applying the recurrence relations for associated Legendre functions 
one could write

\begin{eqnarray}
\Omega _{n,s}\left( d \right) 
& \simeq & \left. -\nabla \left( \phi _{n}^{ i } (x) \right) ^{2} \right|_{x=d/2} \notag \\
&=& - 2 \phi_{n}^{i} \left( x \right) N_{n,s} \left[
\sqrt{\left( 1-x^{2}\right) }P_{s}^{\left( n-s\right) + 1 } + 
\right. ~ \notag \\
&+& \left. \left.
\left( n-s\right) x P_{s}^{\left( n-s\right) } 
\right] \right|_{x= \tanh d /2}
~,  \label{Eq_TunnelRate_ns_PH}
\end{eqnarray}

For $\tanh d /2 \rightarrow 1$
the above expression 
could be further simplified
\begin{equation}
 \Omega _{n,s}\left( d\right) \sim 2\left(\phi _{n,s}^{i}\left( d/2\right) \right)^2 \left( s-n\right) 
= B_{n,s} e^{ -\left( s-n\right) d} ~, \label{Eq_TunnelRate_Asymptotic} 
\end{equation}
where
\begin{equation}
B_{n,s} =\frac{\Gamma \left( 2s-n+1\right) }{\Gamma \left( n+1\right) }\left( \frac{2s\left( s+1\right) }{\Gamma \left( s-n\right) }\right) ^{2}
. \label{Eq_Eq_TunnelRate_Bcoef}
\end{equation}

Therefore,
\begin{equation}
\frac{\Omega _{n,s}\left( d \right) }{\Omega _{n-1,s}\left( d \right) } \sim A_{n,s} e^{d}
~, \label{Eq_TunnelRate_Ratio_Asymptotic} 
\end{equation}
with
\begin{equation}
A_{n,s} =\frac{(s-n)^2}{\left( 2s-n+1\right) n}
~. \label{Eq_TunnelRate_Ratio_Acoef}
\end{equation}

From (9), it is clearly shown that for a fixed potential depth (fixed $s$)
the tunneling rate $\Omega _{n,s}\left( d \right)$ increases with the energy level $n$ and decreases with the distance $d$.
In contrast, the tunneling rate ratio for two consecutive levels, Eq. (11),
exhibits inverse dependence, i.e.
it decreases with the energy level $n$ and increases exponentially with the distance $d$.

\subsection{\label{Conditions} Conditions for two-state filtering}

For the transfer process we will assume that the position of the trap center
for each of the traps can be varied at will to temporally control the tunneling interaction \cite{grangier,latestbirkl}.
In this case, the adiabatic transport will consist in 
approaching and separating the traps in a counterintuitive sequence,
see Fig.~\ref{f1v1}(a), with typical spatial profiles given in Fig.~\ref{f1v1}(b). The two empty traps, right and middle ones,
are approached and separated first and, with an appropriate time delay $T$, 
left and middle traps are approached and separated. 
The motional sequence of the traps, see Fig.~\ref{f1v1}(a), is engineered in such a way
that the time variation of the tunneling rates between two adjacent traps resembles a Gaussian profile, see Fig.~\ref{f1v1}(c).
With this aim and taking into account the explicit dependence of the tunneling rates with the distances, see Eq.~(9), we fix $x_M=0$ and take the following temporal
variation for the outermost trap positions
\begin{eqnarray}
\frac{({x}_{L} - {x}_{M})}{\alpha}
&=&- \sqrt{\omega^2_x\left({t}-\frac{T}{2}\right) ^{2} \left( \frac{v_{0}}{\alpha \omega_{x}} \right)^{2}
+\left(\frac{{d}_{0}}{\alpha}\right)^{2}} \nonumber \\
\frac{({x}_{R} - {x}_{M})}{\alpha}
&=& \sqrt{\omega^2_x\left({t}+\frac{T}{2}\right) ^{2} \left( \frac{v_{0}}{\alpha \omega_{x}} \right)^{2}
+\left(\frac{{d}_{0}}{\alpha}\right)^{2}},  \label{Eq_MWT_HYP_trajectories}
\label{Eq_MWT_HYP_trajectories}
\end{eqnarray}
with ${d}_{0}$ being the 
minimum separation distance between the outermost (either left or right) and the middle trap
achieved at time $t=\pm T/2$, respectively. $v_{0} $ gives the modulus of the velocity of the outermost traps
at large separation distances.

Since we are considering identical traps, the tunneling couplings
between right and middle and between left and middle traps will follow the same dependence with the trap distance,
$\Omega^{LM}_n(d)=\Omega^{MR}_n(d)=\Omega_n(d)$. 
In this case, the ``global'' adiabaticity condition \cite{matterwaveSTIRAP,STIRAP} for spatial adiabatic passage reads:
\begin{eqnarray}
\sqrt{(\Omega^{LM}_n(d_0))^2+(\Omega^{MR}_n(d_0))^2}\,T=\Omega_n(d_0)T>10 
\label{adiabaticity_condition}
\end{eqnarray}
where $T$ is the characteristic time for the adiabatic passage process.

Since we are interested in state-selective atom transfer,
the goal of the filtering protocol will be that the atomic population initially in the highest vibrational state of the left trap follows adiabatically the spatial dark state, $\left|D_1(\theta_1)\right\rangle$ ending in the right trap, while the population initially distributed in the lower levels of the left trap remains there during the whole process.
Therefore, for the two-state filtering case, the counterintuitive motional sequence of the traps should be performed fulfilling:
$\Omega_1 (d_0)T > 10$ and $\Omega_0 (d_0)T \ll 10$. 
Moreover, the filtering protocol requires also to inhibit the direct transfer between neighbouring traps of the population initially in the ground state of the left trap, i.e., $\Omega_0(d_0)T \ll 1$. 
Therefore, the necessary condition for two-state filtering reads:
\begin{equation}
\frac{\Omega_1(d_0)}{\Omega_0(d_0)} \gg 10\,.
\label{filtering_condition}
\end{equation}

Note also that in order to avoid the direct coupling between the outermost traps, 
it is also required that $\Omega_n(2d_0)T \ll 1$ which implies:
\begin{equation}
\frac{\Omega_1(d_0)}{\Omega_1(2d_0)} \gg 10\,.
\label{direct_condition}
\end{equation}

For the filtering sequence shown in Fig.~\ref{f1v1}~(a)-(c) with $\theta_1$ varying from $0$ to $\pi/2$, we have choosen parameters such that both conditions (\ref{filtering_condition}) and (\ref{direct_condition}) are fulfilled and, therefore, one expects that the filtering protocol succeeds.
Figs.~\ref{f1v1}(d) and \ref{f1v1}(e) plot the temporal variation of the population distribution of the left and the right traps, respectively, by integrating the corresponding Schr\"odinger Eq.~(1) with the initial population distribution $P_0^{L}(t=0)=P_1^{L}(t=0)=1/2$. At the end of the process, $P_0^{L}=P_1^{R}=1/2$, which confirms the validity of the filtering protocol.

\begin{figure}[t]
\begin{center}

\includegraphics[scale=1.1,clip=true,angle=0]{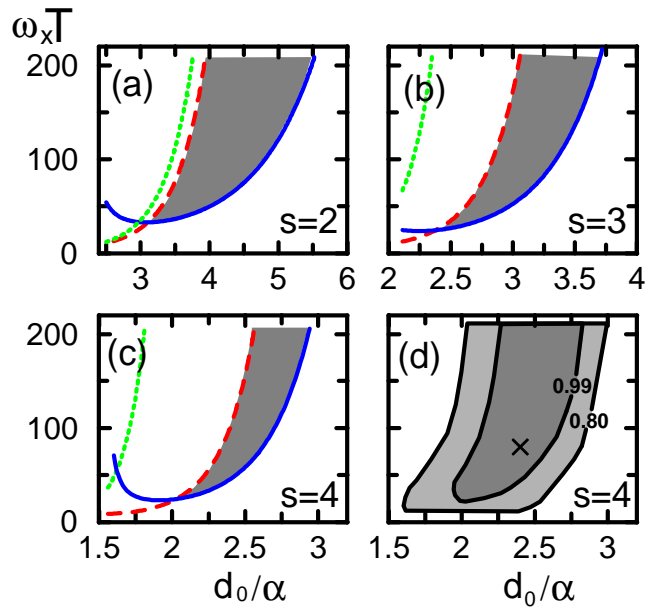}
\end{center}
\caption{
(Color online) Curves $\Omega_1 (d_0) T = 10$ (solid blue) and $\Omega_0 (d_0) T = 1$ (dashed red) in the parameter plane ($d_0/\alpha$, $\omega_xT$) for three different values of the potential depth (a) $s=2$, (b) $s=3$ and (c) $s=4$. In all three cases, the dotted green curve corresponds to $\Omega_1 (2d_0) T=1$. The grey region defines the parameter domain for which the filtering conditions (\ref{filtering_condition}) and (\ref{direct_condition}) are fulfilled. (d) Contour plot of the fidelity (see text) of the filtering process for $s=4$ obtained numerically by integration of the Schr\"{o}dinger equation with the temporal variation of the traps centers given by (\ref{Eq_MWT_HYP_trajectories}).
The cross in (d) marks the parameter setting used in Fig.~\ref{f1v1}.}
\label{f2v1}
\end{figure}

In the following, we will investigate the robustness of the filtering protocol under variations of the parameters. With this aim, we plot in Fig.~\ref{f2v1}(a)-(c) curves $\Omega_1 (d_0) T = 10$ (solid blue) and $\Omega_0 (d_0) T = 1$ (dashed red) in the parameter plane ($d_0/\alpha$, $\omega_xT$) for three different values of the potential depth $s=2, \,3$ and $4$. The dotted green curve corresponds to $\Omega_1 (2d_0) T=1$. The grey region defines the parameter domain for which both conditions (15) and (16) are fulfilled. Note from Fig.~\ref{f2v1} (a)-(c) that even for small values of $s$, the parameter domain where the filtering protocol should succeed is limited by condition (15). To confirm the previous predictions, we have performed numerical simulations of the filtering protocol integrating the Schr\"odinger equation for $s=4$. Fig.~\ref{f2v1}(d) shows the contour plot of the fidelity at the end of the filtering process, defined as $F=P_0^{L}+P_1^{R}$ for an initial population distribution of $P_0^{L}=P_1^{L}=1/2$. For this case, the previously derived filtering conditions (\ref{filtering_condition}) and (\ref{direct_condition}) assure that the fidelity of the process is above 0.99.
We have numerically checked the validity of the derived filtering conditions for a wide set of parameters.

\subsection{\label{multiple} Extension to two excited states }

The generalization of the filtering protocol to two higher excited levels, $n$ and $n-1$, is straightforward. In this case, the counterintuitive motional sequence of the three PT traps should be performed fulfilling conditions $\Omega_{n,s} (d_0)T > 10$ and $\Omega_{n-1,s} (d_0)T \ll 1$, which 
generalizes expression (\ref{filtering_condition}).

\begin{figure}[t]
\begin{center}
\includegraphics[scale=1.0,clip=true,angle=0]{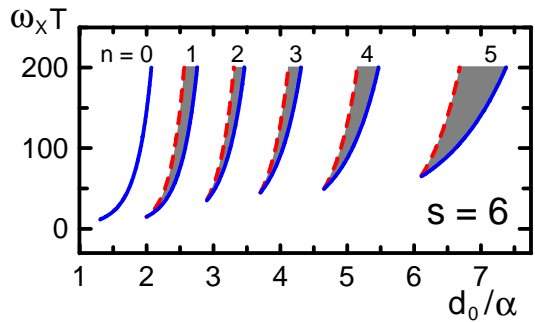}
\end{center}
\caption{
(Color online) Curves $\Omega_{n,s} \left( d_{0} \right) T = 10$ (solid blue) and $\Omega_{n-1,s} \left( d_{0} \right) T = 1$ (dashed red)
in the parameter plane ($d_{0} / \alpha$, $\omega_{x} T$)
for PT potentials with depth $s=6$.
Filtering for the vibrational level $n$ can be achieved in the corresponding grey region.}
\label{f3v1}
\end{figure}

In Fig.~\ref{f3v1} we plot curves $\Omega_{n,s} (d_0) T = 10$ (solid blue) and 
$\Omega_{n-1,s} (d_0) T = 1$ (dashed red) in the parameter plane ($d_0/\alpha$, $\omega_{x} T$) 
for the potential depth $s=6$. Perfect filtering could be performed for each $n$ in the corresponding grey region transferring the population of state $n$ to the right trap without modifying the population of state $n-1$. From Fig.~\ref{f3v1} it is clearly shown that the optimal minimum distance for the filtering protocol decreases with $n$. This minimum distance, $d^{\rm min}_n$, for the filtering protocol involving states $n$ and $n-1$ can be estimated as follows. 
From Eq.~(11) and extending Eq.~(15) to excited states, one obtains:
\begin{equation}
\frac{\Omega _{n,s}\left( d^{\rm min}_n \right) }{\Omega _{n-1,s}\left( d^{\rm min}_n \right) } \sim A_{n,s} e^{d^{\rm min}_n }\gg10, \label{Eq_TunnelRate_Ratio_10} 
\end{equation}
and taking the lower limit of Eq.~(17):
\begin{equation}
d^{\rm min}_n =\ln(\frac{10}{A_{n,s}}).
\label{Eq_d0_estimate} 
\end{equation}
$d^{\rm min}_n$ almost perfectly matches the minimum distance $d_0$ of the lower corner 
of the $n$-th grey region in Fig.~\ref{f3v1}.

\section{\label{multiple} Multiple-state filtering }

In this section, we will extend the previously discussed adiabatic passage technique to the situation where the atomic population is initially distributed among $N+1$ vibrational states of the left PT trap.
We will discuss first a detailed protocol to perform quantum tomography of the atomic population distribution at the left trap and later on we will briefly outline a similar approach for quantum engineering of Fock states. 

\subsection{Quantum Tomography}

To perform quantum tomography, we will apply the filtering protocol 
sequentially in $N$ steps, i.e., state by state, from the most excited 
($n=N$) down to the first excited ($n=1$) state.
At each step $k=\{1,...,N\}$, we will transfer to the right trap 
the population of the corresponding excited state $n=N+1-k$ and 
keeping the rest (from $n-1$ to $0$) in the left trap. 
After each step $k$, the total population in the right trap,
$P^{T}[k]=\sum_n P_n^R [k]$, with $P_n^R [k]$ being the population of state $n$ in the right trap,
will be computed, i.e., measured from the experimental point of view.
After the measurement, the right trap will be emptied and the protocol will be resumed.
After step $k=N$ the population of the left trap is expected to be in its ground vibrational state. Therefore, the last step $k=N+1$ will consist in
directly measuring the total population at the left trap, i.e., $P^{T}[N+1]=\sum_n P_n^L [N]$.
At the end, the set $\{k,P^{T} \left[k \right]\}$ will be the result of the tomography of the initial population distribution at the left trap.
To evaluate the efficiency of the quantum tomography protocol we define the following fidelity:
\begin{equation}
F^{QT}=1-\sum_{n=N}^{0} \left| P_n^L [k=0] - P^{T} [k=N+1-n] \right|,  \label{fidelity}
\end{equation}
where $P_{n}^{L} [k=0]$ is the initial population of state $n$ in the left trap,
while $P^{T} [k=N+1-n]$ is the total population measure at the end of each step $k$. 

Filtering conditions for the step $k$ of the protocol involving states $n$ and $n-1$ read $\Omega_{n,s} (d_0[k])T > 10$ and $\Omega_{n-1,s} (d_0[k])T \ll 1$ implying, as shown in Fig.~\ref{f3v1}, that the minimum distance $d_0[k]$ at each step $k$ should be decreased. Approximated values for $d_0[k]$ at each step of the protocol 
could be estimated by using expression (\ref{Eq_d0_estimate}). However, we will use, in what follows, accurate values of $d_0[k]$ by numerically integrating Eq.~(\ref{Eq_dE_Holstein_Herring}). 

To illustrate the technique outlined above, let us consider three coupled identical 
PT potentials (\ref{Eq_MWT_VxPoschlTeller}) with depth $V_{0}=-156 \hbar \omega_x \left[s=12\right]$ supporting
$12$ bound energy levels. Initially, only the left trap is populated with a truncated thermal distribution among the lowest eight states given by: 
\begin{equation}
P_n^L [k=0]=Z e^{-\frac{E_{n,s}-E_{0,s}}{\beta E_{0,s}}} \quad {\rm with} \, n=0,...,7\,\ \label{Eq_MWT_thermal}
\end{equation}
where $Z$ is the normalization constant and $\beta$ is proportional to the temperature.

\begin{figure}[]
\begin{center}
\includegraphics[scale=0.95,clip=true,angle=0]{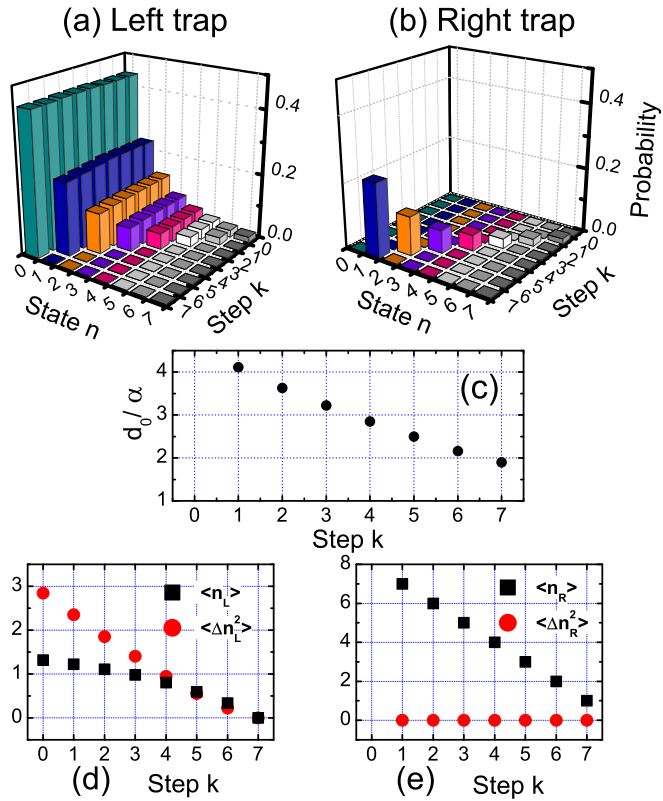}
\end{center}
\caption{
(Color online) 
Quantum tomography protocol via a sequence of adiabatic passage processes
in three PT potentials. Population distribution for the lowest 8 energy levels
in the left (a) and right (b) traps after each step $k$ of the protocol.
The minimum distance between the traps at each step, $d_{0} \left[ k \right]$,
is shown in (c).
Mean $\left< n_i \right>$ (closed squares) and 
variance $\left< \Delta n_i^{2} \right>$ (closed circles)
of the population of each state for the left $(i=L)$ (d) and the right $(i=R)$ (e) traps 
along the protocol. The trap-approaching sequence at each step of the protocol is given by Eq. (\ref{Eq_MWT_HYP_trajectories}) with $\omega_x T/2=40$ and
$v_{0} / \alpha \omega_{x} = 3 \times 10^{-3}$ and the corresponding $d_0[k]$.
}
\label{f8v1}
\end{figure}

Figure \ref{f8v1} shows the distribution of population in the left (a) and the right (b) traps at each step $k$ of the quantum tomography protocol performed with the sequence of distances $d_0[k]$ plotted in (c). The initial distribution is depicted in step $k=0$ of the corresponding figure being the truncated thermal distribution given by Eq.~(\ref{Eq_MWT_thermal}) with $\beta=1$ in the left trap (a) while the right is empty (b). It is clearly seen that at each step $k$, the population of state $n=N+1-k$ is transferred to the right trap while the population of states from $n-1$ to $0$ remains in the left trap. The mean and the variance of the population distribution of the vibrational states of the left and right traps at each step of the protocol are shown in (c) and (d), respectively. Note that the process leads to a sequence of Fock states from $N$ to $1$ at the right trap,
since we assume that after each step the right trap is emptied.
For the left trap, both the mean value of the population distribution and its variance decrease at each step
giving the ground Fock state after the tomography process. The fidelity achieved in the case shown in Fig.~\ref{f8v1} is of $F^{QT}=0.97$.

\subsection{Quantum engineering of Fock states}

\begin{figure}[]
\begin{center}
\includegraphics[scale=0.95,clip=true,angle=0]{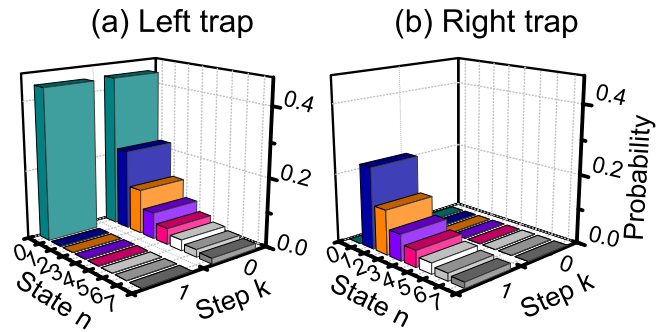}
\end{center}
\caption{
(Color online) Ground state filtering by 
applying a single adiabatic transport process.
Population distribution for the lowest 8 energy levels in the left (a) 
and right (b) traps before and after the single step filtering protocol. 
The minimum distance $d_{0}$ is chosen to fulfill the filtering conditions 
for the first excited vibrational state and it coincides 
with the value used in the last step shown in Fig.\ref{f8v1}. 
The fidelity of the process is above $99\%$.
Other parameters are the same as in Fig. \ref{f8v1}.
}
\label{f10v1}
\end{figure}

The robustness and selectivity of the filtering protocol 
proposed here, allows us to use it for engineering particular Fock states.
We have already seen in the previous subsection that 
it is possible to generate Fock states
at specific excited vibrational levels in the right trap after each step of the tomography protocol.
Moreover, applying the quantum tomography protocol from $k=1$ up to $k=N$,
corresponding to filtering of states from $n=N$ to $n=1$,
one ends up with the Fock state at the ground vibrational level $n=0$ in the left trap. We have to mention that the Fock ground state in the left trap
could be also reached just by one spatial adiabatic passage process
with minimum distance $d_{0}$ chosen to 
perform filtering between the ground and first excited states 
as described in Section \ref{Conditions}.
Under such conditions, all excited states 
will be transferred from the left to the right trap simultaneously, 
since the adiabaticity conditions being fulfilled 
for the first excited vibrational state are fulfilled 
also for all excited states above it.
We have checked numerically such transfer 
obtaining fidelities of the process above $99\%$ (see Fig.\ref{f10v1}).

Alternatively, by combining each of the adiabatic passage steps with a thermalization process it could be possible to implement a tunneling assisted evaporative cooling protocol.
In this case, the protocol would consist in a sequence of 
periodically performed adiabatic spatial passage processes 
but without the need to control precisely the minimum distance $d_{0}$.
In contrast to the standard forced evaporative cooling technique \cite{evapcooling}
developed for magnetic traps, and not easily applicable to dipole traps,
such tunneling assisted (forced) evaporative cooling
could be performed quasi-continuously in dipole traps
without the need to open the trapping potential,
which could be an advantage in 
coherent control and coherent manipulation
of trapped cold atoms and molecules.

\section{\label{Conclusion}Conclusion}

In this paper, we have addressed the filtering of the population of specific vibrational states of a P\"{o}schl--Teller type potential. 
For this purpose, we have chained the initially populated left trap with two empty identical ones and we have performed a vibrational state selective spatial adiabatic passage process from the left trap to the outermost right trap. We have derived analytically the filtering conditions for the two-state case either involving the ground and first excited states as well as two higher excited states and we have applied them to the filtering of an arbitrary number of vibrational states. By numerical integration of the Schr\"odinger equation, we have demonstrated that efficiencies of the protocol above $99$\% can be achieved for a wide set of parameter values leading to the transfer of the population of all vibrational states above a certain one from the left PT trap to the outermost right trap, while the states below it remain at the initial left trap. We have also shown that spatial adiabatic passage 
can be used to perform quantum tomography of the initial population distribution of the left trap with fidelities above $97$\% by applying the filtering protocol starting from the most excited state to the lowest one and sequentially decreasing the minimum distance between the traps at each step. Finally, we have also briefly discussed the possibility of quantum engineering Fock states and of tunneling assisted evaporative cooling.

\begin{acknowledgments}
We acknowledge support from the Spanish Ministry of Science and Innovation under contracts FIS2008-02425, HD2008-0078 and CSD2006-00019 (Consolider project "Quantum Optical Information Technologies"), from the Catalan Government under contract SGR2009-00347
and DAAD (Contract No. 0804149).
\end{acknowledgments}

\appendix

\begin{figure}[]
\begin{center}
\includegraphics[scale=0.95,clip=true,angle=0]{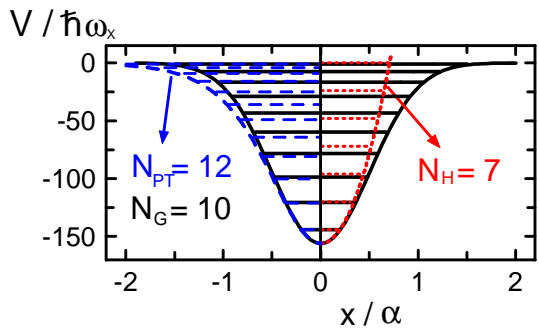}
\end{center}
\caption{
(Color online) 
Gaussian potential (black solid curve) approximated with harmonic (red dotted curve) 
and P\"oschl--Teller (blue dashed curve) potentials.
The corresponding eigenenergies are represented by horizontal black solid (Gaussian potential),
red dotted (harmonic potential) and blue dashed (PT potential) lines.
$N_G$, $N_H$ and $N_{PT}$ give the number of vibrational states 
for the given potential depth for the Gaussian, harmonic and P\"oschl--Teller potentials, respectively.}
\label{f9v1}
\end{figure}

\section{\label{AppendixA}Approximations for Gaussian potentials}

Optical dipole force potentials created by focused light beams
exhibit a Gaussian spatial profile. These type of potentials, due to the lack of analytical 
expressions for their eigenenergies and eigenstates, are very often approximated by other 
types of potentials, typically harmonic ones. In this appendix, we will discuss 
the convenience to use P\"{o}schl--Teller potentials instead of harmonic ones to approximate 
Gaussian potentials. 

A single Gaussian potential in one dimension, black solid line in Fig.~5, can be written as
\begin{eqnarray}
V_{G} \left( x \right) &=& - V_{0} \exp \left( -2 x^{2} / \alpha^{2} \right),
\label{Eq_MWT_1_VxGauss}
\end{eqnarray}
where $V_{0}$ denotes the potential depth and $\alpha=w_{0}$, the waist of the focused light beam. 
The harmonic approximation of this Gaussian potential, Eq.~(A1), reads:
\begin{eqnarray}
V_{H} \left( x \right) &=& - V_{0} \left( 1 -2 x^{2} / \alpha^{2} \right),
\label{Eq_MWT_1_VxHARM}
\end{eqnarray}
and it is depicted in Fig.~5 (red dotted curve). Although the eigenenergies and eigenstates of the harmonic potential can be obtained analytically, it is obvious from Fig.~5 that the harmonic approximation is only accurate for few of the lowest vibrational states. 
Moreover, the harmonic potential has to be cut at some energy value in order to give a finite number of energy eigenstates. 

To account for the finite number of bound states of the Gaussian potentials
it is more convenient to use potentials with known analytical solutions
that support a finite number of bound states. One of such potentials is
the P\"{o}schl--Teller (square hyperbolic secant) potential \cite{PoschlTeller}.
The approximation for the Gaussian potential (\ref{Eq_MWT_1_VxGauss})
with a P\"{o}schl--Teller potential is given by: 
\begin{equation}
V_{PT}(x) = - V_{0} {\rm sech}^{-2} \left( \sqrt{2} x / \alpha \right),
\label{Eq_MWT_1_VxPT}
\end{equation}
corresponding to the blue dashed curve in Fig.~\ref{f9v1}. By comparing the three potentials in Fig.~\ref{f9v1}, it is clear 
that the spectrum of the harmonic potential approximation (red dotted curve) gives six equidistant vibrational states and that its profile fits only close to the bottom of the Gaussian potential. On the contrary, the P\"{o}schl--Teller 
potential spectrum (blue dashed curve) consists of twelve non-equidistant bound vibrational levels and the approximation to the Gaussian potential shape is much more accurate.

\section{\label{AppendixB}Tunneling rate between two P\"{o}schl-Teller potential traps}

The Gram-Schmidt (GS) orthonormalization procedure \cite{jordi}
applied to P\"{o}schl--Teller gives accurate analytical expressions for the tunneling rate of the ground state, $\Omega_{0,s}$.

Considering the single trap eigenstates given by Eq.~(5), the symmetric $\phi _{0,s}^{+}$ and antisymmetric $\phi _{0,s}^{-}$ orthogonal states
for the ground, $n = 0$, vibrational state of two coupled identical PT traps of depth $V_0 = -s \left( s + 1 \right)$ are given by:

\begin{eqnarray}
\phi _{0,s}^{\pm}
&=&
\frac{1}{\sqrt{2}}\left( 
\pm\phi _{0,s} \left( x \right)
 + \phi _{0,s} \left( x-d \right)
\right) = ~ \notag \\
&=&
\frac{
\left( 
\pm ch^{-s}\left( \sqrt{2} \frac{x}{\alpha} \right)
 + ch^{-s}\left( \sqrt{2} \frac{x-d}{\alpha} \right)
\right)
}{\sqrt{2}}
N_{J,0,s}
.  \label{Eq_MWT_PoschlTeller_nS0}
\end{eqnarray}

One could find the energies $E_{0,s}^{\pm } = \left\langle \phi_{0,s}^{\pm} \left| H \right| \phi _{0,s}^{\pm} \right\rangle $
for the eigenstates (\ref{Eq_MWT_PoschlTeller_nS0}) and finally 
the tunneling rate $\Omega _{0,s} (d) = \left| E_{0,s}^{+} (d) - E_{0,s}^{ -} (d) \right|$
for the ground vibrational level of two PT traps separated by a distance 
$d$:

\begin{eqnarray}
\Omega _{0,s} \left( d \right) &=&\frac{2s^{2} W}{A}
\left\{
\frac{\left[ e^{2a}D+e^{-2a}E\right]}{1-W^{2}}
\right.  \notag \\
&-&
\left.
\frac{
\left[
 e^{2a\left( s-1\right) }B
+e^{-2a\left( s-1\right) }C
\right]
W
}{1-W^{2}}
\right\},  \label{Eq_MWT_PoschlTeller_nS0_Rabi_simple}
\end{eqnarray}
with: 
\begin{eqnarray}
\left\langle \phi_{0,s}^{\pm} 
           | \phi_{0,s}^{\pm}
\right\rangle
&=& 1 \pm 
W, \notag
\\
\left\langle 
\phi_{0,s} \left( x\right) |
\phi_{0,s} \left( x-d\right) 
\right\rangle 
&=&
W
= N_{J,0,s}^{2}
\frac{4^{s}}{s}A, \notag
\\
N_{J,0,s}^{2}
&=& 
\frac{\sqrt{2}}{w_{i}}
\frac{s\Gamma \left( s+1/2\right) }{\sqrt{\pi }\Gamma \left( s+1\right) }. \notag
\end{eqnarray}
and
\begin{eqnarray}
A=A\left( s,a\right)  &=&F_{1}\left( s;s,s;s+1;-e^{-2a},-e^{2a}\right) ~,     \notag \\
B=B\left( s,a\right)  &=&F_{1}\left( s+1;2,2s;s+2;-e^{-2a},-e^{2a}\right) ~,  \notag \\
D=D\left( s,a\right)  &=&F_{1}\left( s+1;s,s+2;s+2;-e^{-2a},-e^{2a}\right) ~, \notag
\end{eqnarray}
where $F_{1}\left( \alpha ;\beta ,\beta ^{\prime };\gamma ;z_{1},z_{2}\right) $ is the
Appell hypergeometric function, which is a generalization 
for the hypergeometric functions $_{2}F_{1}\left( \alpha ,\beta ;\gamma ;z\right) $
to two variables $z_{1}$ and $z_{2}$;
$C\left( s,a\right) = B\left( s,-a\right)$ and 
$E\left( s,a\right) = D\left( s,-a\right)$;
$a=d/\sqrt{2}$.

\end{document}